\documentclass[12pt]{article}
\usepackage{amssymb}
\usepackage{color}
\usepackage{indentfirst}
\usepackage{amsmath}
\usepackage{enumerate}
\usepackage{cite}

\newtheorem{theorem}{Theorem}
\newtheorem{lemma}{Lemma}

\newtheorem{example}{Example}

\def\qi#1 {\fbox {\footnote {\ }}\ \footnotetext { From Qi: {\color{red}#1}}}

\begin{document}
\title{Weight distributions of two classes of linear codes with five or six weights}
\author{Xina Zhang\\
College of Mathematics and Statistics, Northwest Normal \\University,
 Lanzhou,  Gansu 730070,  P.R. China\\
Email: zhangxina11@163.com}
\date{}
\maketitle

\begin{abstract}
In this paper, based on the theory of defining sets, two classes of five-weight or six-weight linear codes over $\mathbb{F}_p$ are constructed. The weight distributions of the linear codes are determined by means of Weil sums and a new type of exponential sums. In some case, there is an almost optimal code with respect to Griesmer bound, which is also an optimal one according to the online code table. 


\end{abstract}

\section{Introduction}
In this study, $p$ is an odd prime and assume $q=p^e$ for a positive integer $e$. Let $\mathbb{F}_p$ and $\mathbb{F}_q$ denote the finite field with $p$ and $q$ elements, respectively. We denote by $Tr$ the absolute trace function \cite{LN} from $\mathbb{F}_q$ onto $\mathbb{F}_p$, and use $\mathbb{F}^{*}_q$ and $\mathbb{F}^{*}_p$ to denote the multiplicative group of $\mathbb{F}_q$ and $\mathbb{F}_p$. Obviously, $\mathbb{F}_{q}^{*}=\mathbb{F}_{q}\setminus\{0\}$, and $\mathbb{F}_{p}^{*}=\mathbb{F}_{p}\setminus\{0\}$.

An $[n,k,d]$ \emph{linear code} $\mathcal{C}$ over $\mathbb{F}_q$ is a $k$-dimensional subspace of
$\mathbb{F}_q^n$ with minimum Hamming distance $d$. Let $A_i$ be the number of codewords with Hamming weight $i$ in a code $\mathcal{C}$. The weight enumerator of $\mathcal{C}$ is defined by
$$1+A_1z+A_2z^2+\ldots+A_nz^n,$$
and the sequence $(1, A_1, \ldots, A_n)$ is called the \emph{weight distribution} of $\mathcal{C}$\cite{HP}. If $|\{1\leq i\leq n: A_i\neq 0\}|=t,$ then we say $\mathcal{C}$ a $t$-weight code. In coding theory, the weight distribution of linear codes is an interesting research topic, as it contains important information as to estimate the error correcting capability and the probability of error detection and correction with respect to some algorithms.

Let $\mathcal{C}$ be an $[n,k,d]$ code over $\mathbb{F}_{q}$ with $k\geqslant1$, then the well-known \emph{Griesmer bound}\cite{HP} is given by
$$n\geqslant\sum_{i=0}^{k-1}\lceil\frac{d}{q^{i}}\rceil.$$

An $[n,k,d]$ code is called \emph{optimal} if no $[n,k,d+1]$ code exists, and is called \emph{almost optimal} if the $[n,k,d+1]$ code is optimal\cite{HY}.

One of the constructions of linear codes is based on a proper selection of a subset of finite fields\cite{DCS07}. That is, let $D = \{ {d_1},{d_2}, \ldots ,{d_n}\} \subseteq {F_q}$. A linear code of length $n$ over $\mathbb{F}_p$ is defined as
\begin{equation*}
\mathcal{C}_D =\{(Tr(xd_1),Tr(xd_2), \ldots ,Tr(xd_n)):x \in \mathbb{F}_q\},
\end{equation*}
the set $D$ is called the \emph{defining set} of linear code $\mathcal{C}_D$.  This construction approach is generic in the sense that many classes of optimal linear codes could be produced by selecting the proper defining sets\cite{KD,HY1,TXF,JLF,SY}.

By means of the construction method mentioned above, Zhang et al.\cite{XXRF} constructed a class of linear codes and presented their weight distributions, with the defining set
$D = \{(x_1, x_2) \in \mathbb{F}_q^2 : Tr(x_1^{{p^s} + 1}) = 1, Tr(x_2) = 1  \},$  where $p$ is an odd prime, $q=p^m,$ and $m=2s.$
In this paper, with the same means, we generalize the construction of the defining set, and obtain two classes of linear codes with five or six weights, which include some almost optimal codes. And making use of Weil sums\cite{R,S,C} and a new type of exponential sums, we will determine not only the parameters but also weight distributions of these codes.


\section{Main Results}\label{theorem}
In this section, we present the main results, including the construction, the parameters and the weight distribution of the linear code $\mathcal{C}_{D_i}$. The proofs will be given in the following section.

We begin this section by selecting defining sets
\begin{equation}
D_i = \{(x_1, x_2) \in \mathbb{F}_q^2 : Tr(x_1^{{p^l} + 1}) = 1, Tr(x_2) \in C_i^{(2,p)} \}, i=0,1.
\end{equation}
to construct linear codes
\begin{equation}
\mathcal{C}_{D_i}= \{(Tr(ax_1+bx_2)_{(x_1, x_2)\in D_i}) : a, b \in \mathbb{F}_q  \}, i=0,1.
\end{equation}
where $C_0^{(2,p)}$ and $C_1^{(2,p)}$ are the cyclotomic classes of order $2$ in $\mathbb{F}_p^*$\cite{ST}, also denote the sets of all squares and non-squares in $\mathbb{F}_p^*$, respectively. Let $q=p^e$ satisfying $p \equiv 3 \bmod 4$, and $s=\gcd(l,e)$ be the greatest common divisor of positive integers $l$ and $e$. The code $\mathcal{C}_{D_i}$ will be discussed under the assumption that $e/s$ is even with $e=2m$ and $m\geq1$. Throughout the paper, $\eta$ is the quadratic character over $\mathbb{F}_{p}^{*}$, and is extended by $\eta(0)=0$. Then the weights and weight distributions of the linear codes are studied by utilizing some results of Weil sums\cite{S,C} and a new tool of exponential sums.

The following Theorems \ref{weight1}-\ref{weight2} are the main results of this paper.

\begin{theorem}\label{weight1}
If $m/s \equiv 1 \bmod 2$, then the weight distribution of the codes $\mathcal{C}_{D_i}$\,$(i=0,1)$ with the parameters $[\frac{p-1}{2}(p^{2e-2}+p^{e+m-2}), 2e]$ is listed in table \ref{1}.
Obviously, the codes are at most $6$-weight. When $m=1, s=1$, note that $\frac{p-1}{2}(p^{2e-2}+p^{e+m-2})=\frac{(p-1)^2}{2}p^{2e-3}+(p-1)p^{e+m-2}$, thus the codes are at most $5$-weight.

\begin{table}
\begin{center}
\caption{The weight distribution of $\mathcal{C}_{D_i}$\,($i=0,1$) when $ m/s \equiv 1 \bmod 2 $}\label{1}
\begin{tabular}{ll}
\hline\noalign{\smallskip}
Weight  &  Multiplicity   \\
\noalign{\smallskip}
\hline\noalign{\smallskip}
$0$  &  1 \\
$\frac{p-1}{2}(p^{2e-2}+p^{e+m-2})$  &  $ p-1 $    \\
$\frac{(p-1)^2}{2}(p^{2e-3}+p^{e+m-3})$  &  $p^e(p^e-p)$     \\
$\frac{(p-1)^2}{2}p^{2e-3}$  &  $(p^{e}+p^{e-1}-p^{m}+p^{m-1}-2)/2$     \\
$\frac{(p-1)^2}{2}p^{2e-3}+(p-1)p^{e+m-2}$  &  $(p^{e}-p^{e-1}+p^{m}-p^{m-1})/2$     \\
$\frac{(p-1)^2}{2}p^{2e-3}+\frac{p-1}{2}p^{e+m-2}$ & $(p-1)(p^{e}+p^{e-1}-p^{m}+p^{m-1}-2)/2$     \\
$\frac{(p-1)^2}{2}p^{2e-3}+\frac{p-3}{2}p^{e+m-2}$ & $(p-1)(p^{e}-p^{e-1}+p^{m}-p^{m-1})/2$      \\
\noalign{\smallskip}
\hline
\end{tabular}
\end{center}
\end{table}
\end{theorem}

\begin{theorem}\label{weight2}
If $m \geq s+1$ and $ m/s \equiv 0 \bmod 2$, then the weight distribution of the codes $\mathcal{C}_{D_i}$\,$(i=0,1)$ with the parameters $[\frac{p-1}{2}(p^{2e-2}+p^{e+m+s-2}), 2e]$ is listed in table \ref{2}.
Obviously, the codes are at most $6$-weight. When $m=s+1$,
 note that $\frac{p-1}{2}(p^{2e-2}+p^{e+m+s-2})=\frac{(p-1)^2}{2}p^{2e-3}+(p-1)p^{e+m+s-2}$, thus the codes are at most $5$-weight.

\begin{table}
\begin{center}
\caption{The weight distribution of $\mathcal{C}_{D_i}$\,($i=0,1$) when $ m/s \equiv 0 \bmod 2 $}\label{2}
\begin{tabular}{ll}
\hline\noalign{\smallskip}
Weight  &  Multiplicity   \\
\noalign{\smallskip}
\hline\noalign{\smallskip}
$0$  &  1 \\
$\frac{p-1}{2}(p^{2e-2}+p^{e+m+s-2})$  &  $ p-1 $    \\
$\frac{(p-1)^2}{2}(p^{2e-3}+p^{e+m+s-3})$  &  $p^e(p^e-p^{1-2s})$     \\
$\frac{(p-1)^2}{2}p^{2e-3}$  &  $(p^{e-2s}+p^{e-2s-1}-p^{m-s}+p^{m-s-1}-2)/2$     \\
$\frac{(p-1)^2}{2}p^{2e-3}+(p-1)p^{e+m+s-2}$  &  $(p^{e-2s}-p^{e-2s-1}+p^{m-s}-p^{m-s-1})/2$     \\
$\frac{(p-1)^2}{2}p^{2e-3}+\frac{p-1}{2}p^{e+m+s-2}$ & $ (p-1)(p^{e-2s}+p^{e-2s-1}-p^{m-s}+p^{m-s-1}-2)/2$     \\
$\frac{(p-1)^2}{2}p^{2e-3}+\frac{p-3}{2}p^{e+m+s-2}$ & $ (p-1)(p^{e-2s}-p^{e-2s-1}+p^{m-s}-p^{m-s-1})/2$     \\
\noalign{\smallskip}
\hline
\end{tabular}
\end{center}
\end{table}
\end{theorem}

The followings are some examples about our results verified by Magma.

\begin{example}\label{example1}
If $(p,e,l)=(3, 2, 1)$, then $m=1$, $s=\gcd(e,l)=1$ and $m/s \equiv 1 \bmod 2$. By Theorem \ref{weight1}, the code $\mathcal{C}_{D_i}$\,$(i=0,1)$ has parameters $[12,4,6]$ with weight enumerator $1+12z^{6}+54z^{8}+8z^{9}+6z^{12}$,
which confirmed the result by Magma. According to Griesmer bound, this code is almost optimal as the best linear code of length $12$ and dimension $4$ over $\mathbb{F}_{3}$ has minimum weight $7$. Furthermore, the code is optimal one with respect to the code table\cite{YZ}. In fact, we can get the same result by Magma when $l=3,5,7,9,11$.
\end{example}

\begin{example}\label{example2}
If $(p,e,l)=(3,4,1)$, then $m=2$, $s=\gcd(e,l)=1$ and $m/s \equiv 0 \bmod 2$. By Theorem \ref{weight2}, the code $\mathcal{C}_{D_i}$\,$(i=0,1)$ has parameters $[972,8,486]$ with weight enumerator
$1+12z^{486}+6534z^{648}+8z^{729}+6z^{972}$,
which confirmed the result by Magma. In fact, we can get the same result by Magma when $l=3,5,7,9$.
\end{example}

\begin{example}\label{example3}
If $(p,e,l)=(3, 4, 2)$, then $m=2$, $s=\gcd(e,l)=2$ and $m/s \equiv 1 \bmod 2$. By Theorem \ref{weight1}, the code $\mathcal{C}_{D_i}$\,$(i=0,1)$ has parameters $[810,8,486]$ with weight enumerator $1+110z^{486}+6318z^{540}+100z^{567}+30z^{648}+2z^{810}$,
which confirmed the result by Magma. In fact, we can get the same result by Magma when $l=6,10,14$.
\end{example}

\begin{example}\label{example4}
If $(p,e,l)=(7, 2, 1)$, then $m=1$, $s=\gcd(e,l)=1$ and $m/s \equiv 1 \bmod 2$. By Theorem \ref{weight1}, the code $\mathcal{C}_{D_i}$\,$(i=0,1)$ has parameters $[168,4,126]$ with weight enumerator $1+24z^{126}+144z^{140}+2058z^{144}+144z^{147}+30z^{168}$,
which confirmed the result by Magma. In fact, we can get the same result by Magma when $l=3,5,7,9,11$.
\end{example}

\begin{example}\label{example5}
If $(p,e,l)=(11, 2, 1)$, then $m=1$, $s=\gcd(e,l)=1$ and $m/s \equiv 1 \bmod 2$. By Theorem \ref{weight1}, the code $\mathcal{C}_{D_i}$\,$(i=0,1)$ has parameters $[660,4,550]$ with weight enumerator $1+60z^{550}+600z^{594}+13310z^{600}+600z^{605}+70z^{660}$,
which confirmed the result by Magma. In fact, we can get the same result by Magma when $l=3,5,7,9,11$.
\end{example}

\section{Preliminaries and Auxiliary lemmas}\label{preliminaries}
In this section, we present some facts on exponential sums, that will be needed in calculating the weight enumerator of the codes defined in this article.

An additive character of $\mathbb{F}_q$ is a non-zero function $\chi$ from $\mathbb{F}_q$ to the set of complex numbers of absolute value $1$ such that
$\chi(x+y)=\chi(x)\chi(y)$ for any pair $(x,y) \in \mathbb{F}_q^2$. For each $u \in \mathbb{F}_q$, the function
$$\chi_u(v)=\zeta_{p}^{Tr(uv)},~v \in \mathbb{F}_q$$
denotes an additive character of $\mathbb{F}_q$, where $\zeta_{p}=e^{2\pi i/p}$ is a primitive $p$-th root of unity and $i=\sqrt{-1}$.  Since $\chi_0(v)=1$ for all $v \in \mathbb{F}_q$, which is the trivial additive character of $\mathbb{F}_q$. We call $\chi_1$ the canonical additive character of $\mathbb{F}_q$ and we have $\chi_u(x)=\chi_1(ux)$ for all $u\in\mathbb{F}_q$. The additive character satisfies the orthogonal property \cite{LN}, that is
\begin{eqnarray*}
\sum_{v \in \mathbb{F}_q} \chi_u(v)=\left\{
\begin{array}{ll}
q,   & u=0,\\  
0,    & u\neq0.\\  
\end{array}
\right.
\end{eqnarray*}

Let $h$ be a fixed primitive element of $\mathbb{F}_q$. For each $j=0, 1, \ldots, q-2,$ the function $\lambda_j(h^k)=e^{2\pi ijk/(q-1)}$ for $k=0, 1, \ldots, q-2$ defines a multiplicative character of $\mathbb{F}_q$, we extend these characters by setting $\lambda_j(0)=0$. Let $q$ be odd. For $j=(q-1)/2$ and $v \in \mathbb{F}_{q}^{*}$, we have
\begin{equation*}
  \lambda_{(q-1)/2}(v) =
  \begin{cases}
    1, & \text{if $v$ is the square of an element of $\mathbb{F}_{q}^{*}$,}  \\
    -1, & \text{otherwise,}
  \end{cases}
\end{equation*}
which is called the quadratic character of $\mathbb{F}_{q}$, and is denoted by $\eta'$ in the sequel.
We call $\eta'=\lambda_{(q-1)/2}$ and $\eta=\lambda_{(p-1)/2}$ are the quadratic characters over $\mathbb{F}_q$ and $\mathbb{F}_p$, respectively. The quadratic Gauss sums over $\mathbb{F}_q$ and $\mathbb{F}_p$ are defined respectively by
$$G'(\eta')=\sum\limits_{v \in \mathbb{F}_q}\eta'(v)\chi'_1(v) \quad \mathrm{and} \quad G(\eta)=\sum\limits_{v \in \mathbb{F}_p}\eta(v)\chi_1(v),$$
where $\eta$ and $\chi_1$ are the canonical multiplicative and additive characters of $\mathbb{F}_p$, respectively. Moreover, it is well known that $G'=(-1)^{e-1}\sqrt{p^{*}}^e$ and $G=\sqrt{p^{*}}$, where $p^{*}=\eta(-1)p.$

The following are some basic facts on exponential sums.

\begin{lemma}\label{quadratic sums}(\cite{LN}, Theorem 5.33)
If $f(x)=a_{2}x^{2}+a_{1}x+a_{0} \in \mathbb{F}_{q}[x],$ where $a_{2}\neq 0,$ then
$$\sum_{x \in \mathbb{F}_{q}}\zeta_{p}^{Tr(f(x))}=\zeta_{p}^{Tr(a_{0}-a_{1}^{2}(4a_{2})^{-1})}\eta'(a_{2})G'(\eta'),$$
where $\eta'$ is the quadratic character of $\mathbb{F}_{q}$.
\end{lemma}

\begin{lemma}\label{quadratic character}(\cite{LN}, Theorem 5.48)
With the notation above, we have
\begin{equation*}
  \sum_{x \in \mathbb{F}_{q}}\eta'(f(x)) =
  \begin{cases}
    -\eta'(a_{2}), & a_{1}^{2}-4a_{0}a_{2}\neq 0,\\ 
    (q-1)\eta'(a_{2}), & a_{1}^{2}-4a_{0}a_{2}=0.  
  \end{cases}
\end{equation*}
\end{lemma}

For $\alpha, \beta \in \mathbb{F}_{q}$ and any positive integer $l$, the Weil sums       $S(\alpha, \beta)$ is defined by
$$S(\alpha, \beta)= \sum_{x \in \mathbb{F}_{q}}\zeta_{p}^{Tr(\alpha x^{p^{l}+1}+\beta x)}.$$

We will show some results of $S(\alpha, \beta)$ for $\alpha\neq 0$ and $q$ odd.

\begin{lemma}\label{weil sums1}(\cite{S}, Theorem 2)
Let $s=(l,e)$ and $e/s$ be even with $e=2m$. Then
\begin{equation*}
  S(\alpha, 0) =
  \begin{cases}
   (-1)^{m/s}p^m , & \alpha^{(q-1)/(p^s+1)} \neq (-1)^{m/s},\\   
    (-1)^{m/s+1}p^{m+s}, & \alpha^{(q-1)/(p^s+1)}=(-1)^{m/s}.   
  \end{cases}
\end{equation*}
\end{lemma}

\begin{lemma}\label{weil sums2}(\cite{C}, Theorem 4.7)
Let $\beta \neq 0$ and $e/s$ be even with $e=2m.$ Then $S(\alpha, \beta)=0$ unless the equation $\alpha^{p^{l}}X^{p^{2l}}+\alpha X=-\beta^{p^{l}}$ is solvable. There are two possibilities.
\begin{enumerate}
  \item If $\alpha^{(q-1)/(p^s+1)} \neq (-1)^{m/s},$ then for any choice of $\beta \in \mathbb{F}_q,$ the equation has a unique solution $x_{0}$ and $$S(\alpha, \beta)=(-1)^{m/s}p^{m}\zeta_{p}^{Tr(-\alpha x_{0}^{p^{l}+1})}$$
  \item If $\alpha^{(q-1)/(p^s+1)}=(-1)^{m/s}$ and if the equation is solvable with some solution $x_{0}$, then
  $$S(\alpha, \beta)=(-1)^{m/s+1}p^{m+s}\zeta_{p}^{Tr(-\alpha x_{0}^{p^{l}+1})}$$
\end{enumerate}
\end{lemma}

\begin{lemma}\label{weil sums3}(\cite{S}, Theorem 4.1)
For $e=2m$, the equation $\alpha^{p^{l}} X^{p^{2l}}+\alpha X=0$ is solvable for $X \in \mathbb{F}_{q}^{*}$ if and only if $e/s$ is even and $\alpha^{(q-1)/(p^{s}+1)}=(-1)^{m/s}$. In such cases, there are $p^{2s}-1$ non-zero solutions.
\end{lemma}

There is the fact that $\alpha^{p^{l}}X^{p^{2l}}+\alpha X$ is a permutation polynomial over $\mathbb{F}_{q}$ with $q=p^e$ if and only if $e/s$ is odd or $e/s$ is even with $e=2m$ and $\alpha^{(q-1)/(p^s+1)} \neq (-1)^{m/s}$.

\begin{lemma}\label{weil sums4}(\cite{QFKD})
Let $f(X)=X^{p^{2l}}+X$ and $$S=\{\beta \in \mathbb{F}_{q}: f(X)=-\beta^{p^{l}} is\;solvable\;in\;\mathbb{F}_{q}\}.$$
If $m/s \equiv 0\,mod\,2$, then $|S|=p^{e-2s}$.
\end{lemma}
\textbf{proof:} Take into account that both $e/s$ and $m/s$ are even, we can deduce that $\alpha^{(q-1)/(p^s+1)}=(-1)^{m/s}$.
Then $f(X)=0$ has $p^{2s}$ solutions in $\mathbb{F}_{q}$ from Lemma \ref{weil sums3}. So does the equation $f(X)=-\beta^{p^l}$ with $\beta \in S$. For $\beta_{1}, \beta_{2} \in \mathbb{F}_{q}$ and $\beta_{1} \neq \beta_{2}$, there are no common solutions for equations $f(X)=-\beta_{1}^{p^l}$ and $f(X)=-\beta_{2}^{p^l}$. In addition, for each $\alpha \in \mathbb{F}_{q}$, it is known that $f(\alpha)$ is in $\mathbb{F}_{q}$, there must be some $\beta \in \mathbb{F}_{q}$ such that $f(\alpha)=-\beta^{p^l}$. Since by use of the equation $|S|\cdot p^{2s} =p^e$, we can get the desired conclusion. \hfill$\square$

\begin{lemma}\label{new sums}
Let $p$ an odd prime satisfying $p \equiv 3 \bmod 4$. $\forall x \in \mathbb{F}_p$, $\forall y \in \mathbb{F}_p^*$, $\eta$ is the quadratic character over $\mathbb{F}_p^*$, $C_i^{(2,p)}$\,$(i=0,1)$ are the cyclotomic classes of order $2$ in $\mathbb{F}_p^*$, then
\begin{equation*}
\sum_{x \in C_0^{(2,p)}}\eta(x^2-y)=\sum_{x \in C_1^{(2,p)}}\eta(x^2-y)=
\begin{cases}
0, &  y \in C_0^{(2,p)},\\
-1, &  y \in C_1^{(2,p)}.
\end{cases}
\end{equation*}
\end{lemma}
\textbf{proof:} Since $p \equiv 3 \bmod 4$, $\eta(-1)=-1$. Let $g$ be a primitive element of $\mathbb{F}_p$, that is $\mathbb{F}_p^*=\langle g \rangle$, then
\begin{align*}
C_0^{(2,p)}&=\{g^2,g^4, \ldots, g^{\frac{p+1}{2}}, \ldots, g^{p-1}\},\\
C_1^{(2,p)}&=\{g,g^3, \ldots, g^{\frac{p-1}{2}}, \ldots, g^{p-2}\}.
\end{align*}
Clearly, $(C_0^{(2,p)})^2=(C_1^{(2,p)})^2=C_0^{(2,p)}$, thus \,$\forall y \in \mathbb{F}_p$, we have $$\sum_{x \in C_0^{(2,p)}}\eta(x^2-y)=\sum_{x \in C_1^{(2,p)}}\eta(x^2-y).$$
Let $f_1(x)=x^2-y$, then $\Delta=4y \ne 0$, according to Lemma \ref{quadratic character}, $$\sum_{x \in \mathbb{F}_p}\eta(x^2-y)=-\eta(1)=-1.$$ And $\sum\limits_{x \in \mathbb{F}_p}\eta(x^2-y)=\sum\limits_{x \in C_0^{(2,p)}}\eta(x^2-y)+\sum\limits_{x \in C_1^{(2,p)}}\eta(x^2-y)+\eta(-y)$, so $$\sum\limits_{x \in C_0^{(2,p)}}\eta(x^2-y)+\sum\limits_{x \in C_1^{(2,p)}}\eta(x^2-y)+\eta(-y)=-1.$$
\begin{enumerate}
\item If $y \in C_0^{(2,p)}$, $\eta(-y)=\eta(-1)\eta(y)=-1$, we can deduce $$\sum_{x \in C_0^{(2,p)}}\eta(x^2-y)=\sum_{x \in C_1^{(2,p)}}\eta(x^2-y)=0.$$
\item If $y \in C_1^{(2,p)}$, $\eta(-y)=\eta(-1)\eta(y)=1$, we can have $$\sum_{x \in C_0^{(2,p)}}\eta(x^2-y)=\sum_{x \in C_1^{(2,p)}}\eta(x^2-y)=-1.$$
\end{enumerate}
Then we can get the desired results.\hfill$\Box$

\section{The proofs of the main results}\label{proofs}

The following Lemmas \ref{N(u,v)}-\ref{T2} are essential to determine the lengths                                                                                                                                                                                                                                                                                                                                                                                                                                                             and weight distributions of $\mathcal{C}_{D_i}$\,$(i=0,1)$.

\begin{lemma}\label{N(u,v)}
The length of the code $\mathcal{C}_{D_i}$\,$(i=0,1)$ is
\begin{equation*}
\begin{aligned}
  n_i&=|\{ (x_1,x_2) \in \mathbb{F}_q^2: Tr(x_2) \in C_i^{(2,p)}, Tr(x_1^{p^l + 1}) = 1\}|\\&=
  \begin{cases}
    \frac{p-1}{2}(p^{2e-2}+p^{e+m-2}), & ~\text{$m/s\equiv 1 \bmod 2$,}  \\
    \frac{p-1}{2}(p^{2e-2}+p^{e+m+s-2}), & ~\text{$m/s\equiv 0 \bmod 2$, }
  \end{cases}
  \end{aligned}
\end{equation*}
where $i=0,1$.
\end{lemma}
\textbf{Proof:} By the orthogonal property of additive character, we have
\begin{eqnarray*}
n_i&=&|\{ (x_1,x_2) \in \mathbb{F}_q^2:Tr(x_2) \in C_i^{(2,p)}, Tr(x_1^{p^l + 1}) = 1\}|\\
&=&\sum_{c \in C_i^{(2,p)}}\sum_{x_1, x_2 \in \mathbb{F}_q}(\frac{1}{p} \sum_{y_1 \in \mathbb{F}_p}\zeta_{p}^{y_1(Tr(x_1^{p^l+1}) - 1)} )(\frac{1}{p}\sum_{y_2 \in \mathbb{F}_p} \zeta_{p}^{y_2(Tr(x_2) - c)})\\
&=&p^{-2}\sum_{c \in C_i^{(2,p)}}\sum_{x_1, x_2 \in \mathbb{F}_q}\bigg(1+\sum_{y_1 \in \mathbb{F}_p^*}\zeta_{p}^{y_1(Tr(x_1^{p^l+1}) - 1)}\bigg)\bigg(1+\sum_{y_2 \in \mathbb{F}_p^*}\zeta_{p}^{y_2(Tr(x_2) - c)}\bigg)\\
&=&\frac{p-1}{2}p^{2e - 2} +\Omega_{i1} + \Omega_{i2} + \Omega_{i3},
\end{eqnarray*}
where
\begin{eqnarray*}
\Omega_{i1} &=& p^{-2}\sum_{c \in C_i^{(2,p)}}\sum_{x_1,x_2 \in \mathbb{F}_q}\sum_{y_2 \in \mathbb{F}_p^*}\zeta_{p}^{y_2Tr(x_2)-cy_2}\\
&=& p^{-2}\sum_{c \in C_i^{(2,p)}}\sum_{y _2 \in \mathbb{F}_p^*}\zeta_{p}^{-cy_2} \sum_{x_1,x_2 \in \mathbb{F}_q}\zeta_{p}^{y_2Tr(x_2)}\\
 &=& 0,
\end{eqnarray*}
which is due to the orthogonal property of additive character.
\begin{eqnarray*}
\Omega_{i2} &=& p^{-2}\sum_{c \in C_i^{(2,p)}}\sum_{x_1,x_2\in \mathbb{F}_q} \sum_{y_1 \in \mathbb{F}_p^*} \zeta_{p}^{y_1Tr(x_1^{p^l + 1}) - y_1}\\
&=& p^{-2}\sum_{c \in C_i^{(2,p)}}p^e\sum_{y_1 \in\mathbb{F}_p^*} \zeta_{p}^{- y_1} \sum_{x_1 \in \mathbb{F}_q} \zeta_{p}^{y_1Tr(x_1^{p^l + 1})}\\
&=& p^{e-2}\sum_{c \in C_i^{(2,p)}}\sum_{y_{1} \in \mathbb{F}_p^*} \zeta_{p}^{-y_{1}} \cdot S(y_{1},0)
\end{eqnarray*}
From Lemma \ref{weil sums1}, we have
\begin{equation*}
\Omega_{i2}=
\begin{cases}
  \frac{p-1}{2}p^{e+m-2}, &  m/s \equiv 1 \bmod 2, \\
  \frac{p-1}{2}p^{e+m+s-2}, &  m/s \equiv 0 \bmod 2.
\end{cases}
\end{equation*}
\begin{eqnarray*}
\Omega_{i3}&=& p^{-2}\sum_{c \in C_i^{(2,p)}}\sum_{x_1,x_2\in \mathbb{F}_q}\sum_{y_1 \in \mathbb{F}_p^*}\zeta_{p}^{y_1Tr(x_1^{p^l + 1}) - y_1}\sum_{y_2 \in \mathbb{F}_p^*} \zeta_{p}^{y_2Tr(x_2) - cy_2}\\
&=& p^{-2}\sum_{c \in C_i^{(2,p)}}\sum_{y_1 \in \mathbb{F}_p^*}\zeta_{p}^{ -y_1} \sum_{y_2 \in \mathbb{F}_p^*} \zeta_{p}^{ - cy_2}\sum_{x_1\in \mathbb{F}_q}\zeta_{p}^{y_1Tr(x_1^{p^l + 1})}\sum_{x_2\in \mathbb{F}_q}\zeta_{p}^{y_2Tr(x_2)}\\
&=&0,
\end{eqnarray*}
which is also based on the orthogonal property of additive character.

Then, we can get
\begin{equation*}
  n_i=
  \begin{cases}
    \frac{p-1}{2}(p^{2e-2}+p^{e+m-2}), &  m/s \equiv 1 \bmod 2,\\
    \frac{p-1}{2}(p^{2e-2}+p^{e+m+s-2}), &  m/s \equiv 0 \bmod 2,
  \end{cases}
\end{equation*}
where $i=0,1$. Thus we complete the proof of the lemma.
\hfill$\Box$

For any $a, b\in \mathbb{F}_q$ and any codeword $\mathbf{c}(a, b)\in \mathcal{C}_{D_i}$, to determine the weight enumerators of $\mathcal{C}_{D_i}$\,$(i=0,1)$, let
\begin{equation*}
T_i = |\{ (x_1, x_2) \in \mathbb{F}_q^2:Tr(x_1^{p^l + 1}) = 1,  Tr(x_2) \in C_i^{(2,p)}, Tr(ax_1+bx_2) = 0\}|,~i=0,1.
\end{equation*}
Then it is not difficult to obtain the Hamming weight of $\mathbf{c}(a, b)$, that is
\begin{eqnarray}\label{3}
wt_i(\mathbf{c}(a,b)) = n_i - T_i,~i=0,1.
\end{eqnarray}
From the orthogonal property of additive character again, we have
\begin{eqnarray}
T_i&=&|\{ (x_1, x_2) \in \mathbb{F}_q^2:Tr(x_1^{p^l + 1}) = 1,  Tr(x_2) \in C_i^{(2,p)}, Tr(ax_1+bx_2) = 0\}|\nonumber\\
 &=&\sum_{c \in C_i^{(2,p)}}\sum_{x_1, x_2 \in \mathbb{F}_q}\bigg(\frac{1}{p} \sum_{z_1 \in \mathbb{F}_p} \zeta_{p}^{z_1Tr(x_1^{p^l + 1}) - z_1}\bigg)\bigg(\frac{1}{p}\sum_{z_2 \in \mathbb{F}_p}\zeta_{p}^{z_{2}Tr(x_2)-cz_2}\bigg)\bigg(\frac{1}{p}\sum_{ z_3 \in \mathbb{F}_p} \zeta_{p}^{z_3 Tr(ax_1+bx_2)} \bigg)\nonumber\\
 &=&p^{-3}\sum_{c \in C_i^{(2,p)}}\sum_{x_1, x_2 \in \mathbb{F}_q}\bigg(1+\sum_{z_1 \in \mathbb{F}_p^*} \zeta_{p}^{z_1Tr(x_1^{p^l + 1}) - z_1}\bigg)\bigg(1+\sum_{z_2 \in \mathbb{F}_p^*}\zeta_{p}^{z_{2}Tr(x_2)-cz_2}\bigg)\bigg(1+\sum_{ z_3 \in \mathbb{F}_p^*} \zeta_{p}^{z_3 Tr(ax_1+bx_2)} \bigg)\nonumber\\
 &=&\frac{n_i}{p}+ \psi_{i1} + \psi_{i2}+ \psi_{i3}+ \psi_{i4},\label{5}
\end{eqnarray}
where $i=0,1$. Next, we will evaluate $\psi_{i1}$, $\psi_{i2}$, $\psi_{i3}$ and $\psi_{i4}$, separately.
\begin{eqnarray*}
\psi_{i1} &=& p^{-3}\sum_{c \in C_i^{(2,p)}}\sum_{x_1, x_2 \in \mathbb{F}_q}\sum_{ z_3 \in \mathbb{F}_p^*} \zeta_{p}^{z_3 Tr(ax_1+bx_2)}\\
&=& p^{-3}\sum_{c \in C_i^{(2,p)}}\sum_{z_3  \in \mathbb{F}_p^*} \sum_{x_1 \in \mathbb{F}_q}\zeta_{p}^{Tr(az_3x_1)}\sum_{x_2 \in \mathbb{F}_q}\zeta_{p}^{Tr(bz_3x_2)}\\
&=& \left\{
\begin{array}{ll}
\frac{(p - 1)^2}{2}p^{2e-3}, & a=b=0, \\
0, & \text{otherwise},
\end{array}
\right.
\quad\text{where}~i=0,1,
\end{eqnarray*}
which is by means of the orthogonal property of additive character.
\begin{eqnarray*}
\psi_{i2} &=& p^{-3}\sum_{c \in C_i^{(2,p)}}\sum_{x_1, x_2 \in \mathbb{F}_q}\sum_{z_2 \in \mathbb{F}_p^*}\zeta_{p}^{z_{2}Tr(x_2)-cz_2}\sum_{ z_3 \in \mathbb{F}_p^*} \zeta_{p}^{z_3 Tr(ax_1+bx_2)}\\
&=& p^{-3}\sum_{c \in C_i^{(2,p)}}\sum_{z_2 \in \mathbb{F}_p^*} \zeta_{p}^{-cz_2} \sum_{z_3  \in \mathbb{F}_p^*} \sum_{x_1\in \mathbb{F}_q}\zeta_{p}^{ Tr(az_3x_1)}\sum_{x_2\in \mathbb{F}_q}\zeta_{p}^{ Tr((z_2+bz_3)x_2)}
\end{eqnarray*}
with the orthogonal property of additive character again,
\begin{eqnarray*}
\psi_{i2}&=& \left\{
\begin{array}{ll}
p^{e-3}\sum\limits_{c \in C_i^{(2,p)}}\sum\limits_{z_2\in\mathbb{F}_p^*}\zeta_{p}^{-cz_2} \sum\limits_{z_3  \in \mathbb{F}_p^*}\sum\limits_{x_2\in \mathbb{F}_q} \zeta_{p}^{ Tr((z_2+bz_3)x_2)}, & a = 0,\\
  0,  & a \neq 0,
\end{array}
\right.\\
&=& \left\{
\begin{array}{ll}
p^{2e-3}\sum\limits_{c \in C_i^{(2,p)}}\sum\limits_{z_2\in\mathbb{F}_p^*}\zeta_{p}^{-cz_2}\\+ p^{e-3}\sum\limits_{c \in C_i^{(2,p)}}\sum\limits_{z_2\in\mathbb{F}_p^*}\zeta_{p}^{-cz_2}\sum\limits_{z_3 \neq -b^{-1}z_2}
\sum\limits_{x_2\in \mathbb{F}_q}\zeta_{p}^{Tr((z_2+bz_3)x_2)}, & a = 0,~b \in \mathbb{F}_p^*,\\
  0,  & \text{otherwise},
\end{array}
\right.\\
&=& \left\{
\begin{array}{ll}
-\frac{p-1}{2}{p^{2e-3}} , &a =0,~b \in \mathbb{F}_p^*, \\
  0, &\text{otherwise},
\end{array}
\right.
\quad \text{where}~i=0,1.
\end{eqnarray*}
\begin{eqnarray*}
\psi_{i3} &=& p^{-3}\sum_{c \in C_i^{(2,p)}}\sum_{x_1,x_2 \in \mathbb{F}_q}\sum_{z_1 \in \mathbb{F}_p^*}\zeta_{p} ^{z_1(Tr(x_1^{p^l + 1})-1)}\sum_{z_3  \in \mathbb{F}_p^*}\zeta_{p}^{z_3Tr(ax_1+bx_2)}\\
&=& p^{-3}\sum_{c \in C_i^{(2,p)}}\sum_{z_1 \in \mathbb{F}_p^*} \zeta_{p} ^{-z_1}\sum_{z_3 \in \mathbb{F}_p^*}\sum_{x_1\in \mathbb{F}_q}\zeta_{p}^{Tr(z_1x_1^{p^l+1}+az_3x_1)}\sum_{x_2\in \mathbb{F}_q}\zeta_{p} ^{Tr(bz_3x_2)}
\end{eqnarray*}
also utilizing the orthogonal property of additive character, we can have
\begin{equation*}
\psi_{i3}= \left\{
\begin{array}{ll}
\frac{p-1}{2}p^{e-3}\sum\limits_{z_1 \in \mathbb{F}_p^*}\zeta_{p}^{-z_1}\sum\limits_{z_3  \in \mathbb{F}_p^*}\sum\limits_{x_1\in \mathbb{F}_q}\zeta_{p}^{Tr(z_1x_1^{p^l+1}+az_3x_1)}, &  b = 0, \\
0, &  b \ne 0.
\end{array}
\right.\\
\end{equation*}
\indent We will give the specific values of $\psi_{i3}$ under the three conditions in the form of lemma below.

\begin{lemma}\label{W1}
With the notations above, if $b=0$ and $a=0$, we have
\begin{equation*}
\psi_{i3}=
\begin{cases}
\frac{(p-1)^2}{2}p^{e+m-3}, &  m/s \equiv 1 \bmod 2,\\
\frac{(p-1)^2}{2}p^{e+m+s-3}, &  m/s \equiv 0 \bmod 2,
\end{cases}
\quad \text{where}~i=0,1.
\end{equation*}
\end{lemma}
\textbf{Proof:}  If $a=0$, with the definition of Weil sums,
\begin{equation*}
\sum\limits_{x_1 \in \mathbb{F}_{q}}\zeta_{p}^{Tr(z_1 x_1^{p^l+1}+az_3x_1)}=S(z_1, 0).
\end{equation*}
\begin{enumerate}
  \item If $m/s \equiv 1 \bmod 2$, by Lemma \ref{weil sums1}: $$S(z_1, 0)=-p^{m}.$$
      Then we can have
      \begin{displaymath}
        \begin{aligned}
        \psi_{i3}&=\frac{p-1}{2}p^{e-3}\sum\limits_{z_1 \in \mathbb{F}_p^*}\zeta_{p}^{-z_1}\sum\limits_{z_3  \in \mathbb{F}_p^*}S(z_1,0)\\
        &=\frac{p-1}{2}p^{e-3} \cdot (-1) \cdot (p-1) \cdot (-p^{m})\\
        &=\frac{(p-1)^2}{2}p^{e+m-3},
        \end{aligned}
      \end{displaymath}
  \item If $m/s \equiv 0 \bmod 2$, by Lemma \ref{weil sums1}: $$S(z_1, 0)=-p^{m+s}.$$
      Then
      \begin{displaymath}
        \begin{aligned}
        \psi_{i3}&=\frac{p-1}{2}p^{e-3}\sum\limits_{z_1 \in \mathbb{F}_p^*}\zeta_{p}^{-z_1}\sum\limits_{z_3  \in \mathbb{F}_p^*}S(z_1,0)\\
        &=\frac{p-1}{2}p^{e-3} \cdot (-1) \cdot (p-1) \cdot (-p^{m+s})\\
        &=\frac{(p-1)^2}{2}p^{e+m+s-3},
        \end{aligned}
      \end{displaymath}
\end{enumerate}
where $i=0,1$, which lead to the desired conclusion.\hfill$\Box$

\begin{lemma}\label{W2}
With the notations above, if $b=0$, $a \neq 0$, $m/s \equiv 1 \bmod 2$, then the equation $X^{p^{2l}}+X=-a^{p^l}$
has a unique solution $\gamma$ in $\mathbb{F}_{q}$, and
\begin{equation*}
\psi_{i3}=
\begin{cases}
-\frac{p-1}{2}p^{e+m-3}-\frac{p-1}{2}p^{e+m-2}\eta(-Tr(\gamma^{p^l+1})), &  Tr(\gamma^{p^l+1}) \neq 0,\\
\frac{(p-1)^2}{2}p^{e+m-3}, &  Tr(\gamma^{p^l+1}) = 0,
\end{cases}
\quad \text{where}~i=0,1.
\end{equation*}
\end{lemma}
\textbf{Proof:} If $m/s \equiv 1 \bmod 2$, since $\alpha=1$, we have $\alpha^{\frac{q-1}{p^s+1}} \neq (-1)^{m/s}$. By use of Lemma \ref{weil sums3}, it is clearly that $X^{p^{2l}}+X$ is a permutation polynomial over $\mathbb{F}_q$ and $X^{p^{2l}}+X=-a^{p^l}$ has a unique solution $\gamma$ in $\mathbb{F}_{q}$. And $z_3z_1^{-1}\gamma$ is the unique solution of the equation $(z_1X)^{p^{2l}}+z_1X=-(az_3)^{p^{l}}$ for any $z_1,z_3 \in \mathbb{F}_{p}^*$. By Lemma \ref{weil sums2},
\begin{displaymath}
\begin{aligned}
  S(z_1, az_3)&=(-1)^{m/s}p^m \zeta_{p}^{Tr(-z_1(z_3z_1^{-1}\gamma)^{p^l+1})}\\
              &=
              \begin{cases}
                -p^m \zeta_{p}^{-\frac{z_3^2}{z_1}Tr(\gamma^{p^l+1})}, & Tr(\gamma^{p^l+1}) \neq 0, \\
                -p^m, & Tr(\gamma^{p^l+1})=0.
              \end{cases}
\end{aligned}
\end{displaymath}

\begin{enumerate}
  \item If $Tr(\gamma^{p^l+1}) \neq 0$,
   \begin{align*}
      \psi_{i3}&=\frac{p-1}{2}p^{e-3}\sum\limits_{z_1 \in \mathbb{F}_{p}^{*}}\zeta_{p}^{-z_1}\sum\limits_{z_3 \in \mathbb{F}_{p}^{*}}S(z_1, az_3)\\
                &=-\frac{p-1}{2}p^{e+m-3}\sum\limits_{z_1 \in \mathbb{F}_{p}^{*}}\zeta_{p}^{-z_1}\sum\limits_{z_3 \in \mathbb{F}_{p}^{*}}\zeta_{p}^{-\frac{z_3^2}{z_1}Tr(\gamma^{p^l+1})}\\
                &=-\frac{p-1}{2}p^{e+m-3}\sum\limits_{z_1 \in \mathbb{F}_{p}^{*}}\zeta_{p}^{-z_1}\bigg(\sum\limits_{z_3 \in \mathbb{F}_{p}}\zeta_{p}^{-\frac{z_3^2}{z_1}Tr(\gamma^{p^l+1})}-1\bigg)\\
                &=-\frac{p-1}{2}p^{e+m-3}\sum\limits_{z_1 \in \mathbb{F}_{p}^{*}}\zeta_{p}^{-z_1}\sum\limits_{z_3 \in \mathbb{F}_{p}}\zeta_{p}^{-\frac{z_3^2}{z_1}Tr(\gamma^{p^l+1})}+\frac{p-1}{2}p^{e+m-3}\sum\limits_{z_1 \in \mathbb{F}_{p}^{*}}\zeta_{p}^{-z_1}
   \end{align*}
   by use of Lemma \ref{quadratic sums} and the orthogonal property of additive character, we have
   \begin{equation*}
      \psi_{i3}=-\frac{p-1}{2}p^{e+m-3}\sum\limits_{z_1 \in \mathbb{F}_{p}^{*}}\zeta_{p}^{-z_1}\eta\Big(-\frac{1}{z_1}Tr(\gamma^{p^l+1})\Big)G(\eta)-\frac{p-1}{2}p^{e+m-3}
   \end{equation*}
   with the definition of quadratic Gauss sums over $\mathbb{F}_p$, we have
   \begin{align*}
      \psi_{i3}&=-\frac{p-1}{2}p^{e+m-3}(G(\eta))^2\eta\big(Tr(\gamma^{p^l+1})\big)-\frac{p-1}{2}p^{e+m-3}\\
               &=-\frac{p-1}{2}p^{e+m-2}\eta\big(-Tr(\gamma^{p^l+1})\big)-\frac{p-1}{2}p^{e+m-3},
   \end{align*}
  \item If $Tr(\gamma^{p^l+1}) = 0$,
    \begin{align*}
     \psi_{i3}&=\frac{p-1}{2}p^{e-3}\sum\limits_{z_1 \in \mathbb{F}_{p}^{*}}\zeta_{p}^{-z_1}\sum\limits_{z_3 \in \mathbb{F}_{p}^{*}}S(z_1, az_3)\\
              &=\frac{p-1}{2}p^{e-3}\sum\limits_{z_1 \in \mathbb{F}_{p}^{*}}\zeta_{p}^{-z_1}\sum\limits_{z_3 \in \mathbb{F}_{p}^{*}}(-p^m)\\
               &=\frac{p-1}{2}p^{e-3} \cdot (-1) \cdot (p-1) \cdot (-p^m)\\
              &=\frac{(p-1)^2}{2}p^{e+m-3},
    \end{align*}
    where the result is derived from the orthogonal property of the additive character.
\end{enumerate}
where $i=0,1$. Then we can get the desired conclusion.\hfill$\Box$

\begin{lemma}\label{W3}
With the notations above, let $b=0$, $a \neq 0$, $m/s \equiv 0 \bmod 2$. If the equation $X^{p^{2l}}+X=-a^{p^l}$
has no solution in $\mathbb{F}_{q}$, then $\psi_{i3}=0, i=0,1$. Suppose that $X^{p^{2l}}+X=-a^{p^l}$
has a solution $\gamma$ in $\mathbb{F}_{q}$, we have
\begin{equation*}
\psi_{i3}=
\begin{cases}
-\frac{p-1}{2}p^{e+m+s-3}-\frac{p-1}{2}p^{e+m+s-2}\eta(-Tr(\gamma^{p^l+1})), &  Tr(\gamma^{p^l+1}) \neq 0,\\
\frac{(p-1)^2}{2}p^{e+m+s-3}, &  Tr(\gamma^{p^l+1}) = 0,
\end{cases}
\quad \text{where}~i=0,1.
\end{equation*}
\end{lemma}
\textbf{Proof:} If $m/s \equiv 0 \bmod 2$, since $\alpha=1$, we have $\alpha^{\frac{q-1}{p^s+1}} = (-1)^{m/s}$. By Lemma \ref{weil sums2}, $S(z_1, az_3)=0$ unless the equation $X^{p^{2l}}+X=-a^{p^l}$ is solvable. From Lemma \ref{weil sums3}, $f(X)=X^{p^{2l}}+X$ is not permutation polynomial over $\mathbb{F}_q$. \\
\indent If the equation $X^{p^{2l}}+X=-a^{p^l}$
has no solution in $\mathbb{F}_{q}$, then $S(z_1, az_3)=0$, thus $\psi_{i3}=0, i=0,1$. \\
\indent If the equation $X^{p^{2l}}+X=-a^{p^l}$
has a solution $\gamma$ in $\mathbb{F}_{q}$, with Lemma \ref{weil sums2},
\begin{displaymath}
\begin{aligned}
  S(z_1, az_3)&=(-1)^{m/s+1}p^{m+s} \zeta_{p}^{Tr(-z_1(z_3z_1^{-1}\gamma)^{p^l+1})}\\
              &=(-1)^{m/s+1}p^{m+s} \zeta_{p}^{-\frac{z_3^2}{z_1}Tr(\gamma^{p^l+1})}\\&=
              \begin{cases}
                -p^{m+s} \zeta_{p}^{-\frac{z_3^2}{z_1}Tr(\gamma^{p^l+1})}, & Tr(\gamma^{p^l+1}) \neq 0, \\  
                -p^{m+s}, & Tr(\gamma^{p^l+1})=0.  
              \end{cases}
\end{aligned}
\end{displaymath}
The following proof process is similar to that of Lemma \ref{W2}, and will not be included here.\hfill$\square$
\begin{eqnarray*}
\psi_{i4}& =& p^{-3}\sum_{c \in C_i^{(2,p)}}\sum_{x_1,x_2 \in \mathbb{F}_q} \sum_{z_1 \in \mathbb{F}_p^*} \zeta_{p}^{z_1(Tr(x_1^{p^l+1} - 1)} \sum_{z_2 \in \mathbb{F}_p^*} \zeta_{p}^{z_2(Tr(x_2)-c)} \sum_{z_3  \in \mathbb{F}_p^*} \zeta_{p}^{z_3Tr(ax_1+bx_2)}\\
&=&p^{-3}\sum_{c \in C_i^{(2,p)}}\sum_{z_1 \in \mathbb{F}_p^*} \zeta_{p}^{-z_1}\sum_{z_2 \in \mathbb{F}_p^*} \zeta_{p}^{-cz_2}\sum_{z_3 \in \mathbb{F}_p^*}\sum_{x_1 \in \mathbb{F}_q}\zeta_{p}^{Tr(z_1x_1^{p^l+1}+az_3x_1)}\sum_{x_2 \in \mathbb{F}_q}\zeta_{p}^{Tr((z_2+bz_3)x_2)}\\
\end{eqnarray*}

\begin{enumerate}
  \item If $b \in \mathbb{F}_{q}/\mathbb{F}_{p}^*$, we have $\psi_{i4}=0,\,i=0,1$.
  \item If $b \in \mathbb{F}_{p}^*$, we have
  \begin{align*}
    \psi_{i4}&=p^{-3}\sum_{c \in C_i^{(2,p)}}\sum_{z_1 \in \mathbb{F}_{p}^*}\zeta_{p}^{-z_1}\sum_{z_2 \in \mathbb{F}_{p}^*}\zeta_{p}^{-cz_2}(\sum_{x_1 \in \mathbb{F}_{q}}\zeta_{p}^{Tr(z_1x_1^{p^l+1}+a(-b^{-1}z_2)x_1)}\sum_{x_2 \in \mathbb{F}_q}\zeta_{p}^0\\
    &+\sum_{z_3 \neq -b^{-1}z_2}\sum_{x_1 \in \mathbb{F}_q}\zeta_{p}^{Tr(z_1x_1^{p^l+1}+az_3x_1)}\sum_{x_2 \in \mathbb{F}_q}\zeta_{p}^{Tr((z_2+bz_3)x_2)})\\
    &=p^{e-3}\sum_{c \in C_i^{(2,p)}}\sum_{z_1 \in \mathbb{F}_p^*}\zeta_{p}^{-z_1}\sum_{z_2 \in \mathbb{F}_p^*}\zeta_{p}^{-cz_2}\sum_{x_1 \in \mathbb{F}_q}\zeta_{p}^{Tr(z_1x_1^{p^l+1}+(-ab^{-1}z_2)x_1)}\\
    &=p^{e-3}\sum_{c \in C_i^{(2,p)}}\sum_{z_1 \in \mathbb{F}_p^*}\zeta_{p}^{-z_1}\sum_{z_2 \in \mathbb{F}_p^*}\zeta_{p}^{-cz_2}S(z_1, -ab^{-1}z_2),
    \end{align*}
    where $i=0,1$.
\end{enumerate}

In the sequel, we will give the rest values of $\psi_{i4}$ through the other three lemmas.

\begin{lemma}\label{W4}
With the notations above, if $b \in \mathbb{F}_p^*$ and $a=0$, we have
\begin{equation*}
\psi_{i4}=
\begin{cases}
-\frac{p-1}{2}p^{e+m-3}, &  m/s \equiv 1 \bmod 2,\\
-\frac{p-1}{2}p^{e+m+s-3}, &  m/s \equiv 0 \bmod 2,
\end{cases}
\quad \text{where}~i=0,1.
\end{equation*}
\end{lemma}
\textbf{Proof:} In this case, with Lemma \ref{weil sums1} and the orthogonal property of the additive character, the proof of this lemma is analogous to that in Lemma \ref{W1}. The details are omitted.\hfill$\square$

\begin{lemma}\label{W5}
With the notations above, if $b \in \mathbb{F}_{p}^*$, $a \neq 0$, $m/s \equiv 1 \bmod 2$, then the equation $X^{p^{2l}}+X=-a^{p^l}$
has a unique solution $\gamma$ in $\mathbb{F}_{q}$, and
\begin{equation*}
\psi_{i4}=
\begin{cases}
-\frac{p-1}{2}p^{e+m-3}, &  Tr(\gamma^{p^l+1}) = 0,\\
-\frac{p-1}{2}p^{e+m-3}, &  Tr(\gamma^{p^l+1}) \in C_0^{(2,p)},\\
\frac{p+1}{2}p^{e+m-3}, &  Tr(\gamma^{p^l+1}) \in C_1^{(2,p)},
\end{cases}
\quad \text{where}~i=0,1.
\end{equation*}
\end{lemma}
\textbf{Proof:} With Lemma \ref{weil sums2}, by means of the similar proof process of Lemma \ref{W2}, the proof process is as follows:
\begin{enumerate}
\item If $Tr(\gamma^{p^l+1})=0$,
     \begin{align*}
       \psi_{i4}&=p^{e-3}\sum_{c \in C_i^{(2,p)}}\sum_{z_1 \in \mathbb{F}_p^*}\zeta_{p}^{-z_1}\sum_{z_2 \in \mathbb{F}_p^*}\zeta_{p}^{-cz_2}S(z_1, -ab^{-1}z_2)\\
              &=p^{e-3}\sum_{c \in C_i^{(2,p)}}\sum\limits_{z_1 \in \mathbb{F}_{p}^{*}}\zeta_{p}^{-z_1}\sum\limits_{z_2 \in \mathbb{F}_{p}^{*}}\zeta_{p}^{-cz_2} \cdot (-p^m)\\
              &=-\frac{p-1}{2}p^{e+m-3},
     \end{align*}
\item If $Tr(\gamma^{p^l+1}) \ne 0$,
      \begin{eqnarray*}
       \psi_{i4}&=&p^{e-3}\sum_{c \in C_i^{(2,p)}}\sum_{z_1 \in \mathbb{F}_p^*}\zeta_{p}^{-z_1}\sum_{z_2 \in \mathbb{F}_p^*}\zeta_{p}^{-cz_2}S(z_1, -ab^{-1}z_2)\\
             &=&p^{e-3}\sum_{c \in C_i^{(2,p)}}\sum_{z_1 \in \mathbb{F}_p^*}\zeta_{p}^{-z_1}\sum_{z_2 \in \mathbb{F}_p^*}\zeta_{p}^{-cz_2}\bigg(-p^m\zeta_{p}^{-\frac{z_2^2}{b^2z_1}Tr(\gamma^{p^l+1})}\bigg)\\
             &=&-p^{e+m-3}\sum_{c \in C_i^{(2,p)}}\sum\limits_{z_1 \in \mathbb{F}_{p}^{*}}\zeta_{p}^{-z_1}\bigg(\sum\limits_{z_2 \in \mathbb{F}_{p}}\zeta_{p}^{-\frac{Tr(\gamma^{p^l+1})}{b^2z_1}z_2^2-cz_2}-1\bigg)\\
             &=&-p^{e+m-3}\sum_{c \in C_i^{(2,p)}}\sum\limits_{z_1 \in \mathbb{F}_{p}^{*}}\zeta_{p}^{-z_1}\sum\limits_{z_2 \in \mathbb{F}_{p}}\zeta_{p}^{-\frac{Tr(\gamma^{p^l+1})}{b^2z_1}z_2^2-cz_2}+\frac{p-1}{2}p^{e+m-3}\sum\limits_{z_1 \in \mathbb{F}_{p}^{*}}\zeta_{p}^{-z_1}
       \end{eqnarray*}
       with Lemma \ref{quadratic sums} and the orthogonal property of the additive character,
       \begin{eqnarray*}
       \psi_{i4}&=&-p^{e+m-3}\sum_{c \in C_i^{(2,p)}}\sum\limits_{z_1 \in \mathbb{F}_{p}^{*}}\zeta_{p}^{-z_1}\zeta_{p}^{-\frac{(-c)^2}{4\big(-\frac{1}{b^2z_1}Tr(\gamma^{p^l+1})\big)}}
                \eta\Big(-\frac{1}{b^2z_1}Tr(\gamma^{p^l+1})\Big)G(\eta)-\frac{p-1}{2}p^{e+m-3}\\
             &=&-p^{e+m-3}\sum_{c \in C_i^{(2,p)}}\sum\limits_{z_1 \in \mathbb{F}_{p}^{*}}\zeta_{p}^{\frac{(bc)^2-4Tr(\gamma^{p^l+1})}{4Tr(\gamma^{p^l+1})}z_1}
             \eta\bigg(\frac{(bc)^2-4Tr(\gamma^{p^l+1})}{4Tr(\gamma^{p^l+1})}z_1\bigg)\eta(-1)G(\eta)\\
             & &\eta\big((bc)^2-4Tr(\gamma^{p^l+1})\big)-\frac{p-1}{2}p^{e+m-3}
       \end{eqnarray*}
       by the definition of the quadratic Gauss sums over $\mathbb{F}_p$,
       \begin{align*}
          \psi_{i4}&=-p^{e+m-3}(G(\eta))^2 \cdot \eta(-1)\cdot\sum_{c \in C_i^{(2,p)}}\eta\big((bc)^2-4Tr(\gamma^{p^l+1})\big)-\frac{p-1}{2}p^{e+m-3}\\
          &=-p^{e+m-2}\sum_{c \in C_i^{(2,p)}}\eta\big((bc)^2-4Tr(\gamma^{p^l+1})\big)-\frac{p-1}{2}p^{e+m-3}
      \end{align*}
      by using Lemma \ref{quadratic character},
      \begin{eqnarray*}
      \psi_{i4}&=&-p^{e+m-2}\cdot \frac{1}{2}\big(\sum_{c \in \mathbb{F}_p}\eta(b^2c^2-4Tr(\gamma^{p^l+1}))-\eta(-Tr(\gamma^{p^l+1}))\big)-\frac{p-1}{2}p^{e+m-3}\\
      &=&-p^{e+m-2}\cdot \frac{1}{2}\big(-1-\eta(-Tr(\gamma^{p^l+1}))\big)-\frac{p-1}{2}p^{e+m-3}
      \end{eqnarray*}
      If $p \equiv 3 \bmod 4$, $\eta(-1)=(-1)^{\frac{p-1}{2}}=-1$, that is $-1 \in C_1^{(2,p)}$, with Lemma \ref{new sums}, we have
      \begin{equation*}
      \psi_{i4}=
      \begin{cases}
      -\frac{p-1}{2}p^{e+m-3}, & Tr(\gamma^{p^l+1}) \in C_0^{(2,p)},\\
      \frac{p+1}{2}p^{e+m-3}, & Tr(\gamma^{p^l+1}) \in C_1^{(2,p)},
      \end{cases}
      \end{equation*}
\end{enumerate}
where $i=0,1$. Then we deduce the desired result.\hfill$\Box$

\begin{lemma}\label{W6}
With the notations above, let $b \in \mathbb{F}_p^*$, $a \neq 0$, $m/s \equiv 0 \, mod \, 2$. If the equation $X^{p^{2l}}+X=-a^{p^l}$
has no solution in $\mathbb{F}_{q}$, then $\psi_{i4}=0, i=0,1$. Suppose that $X^{p^{2l}}+X=-a^{p^l}$
has a solution $\gamma$ in $\mathbb{F}_{q}$, we have
\begin{equation*}
\psi_{i4}=
\begin{cases}
-\frac{p-1}{2}p^{e+m+s-3}, &  Tr(\gamma^{p^l+1}) = 0,\\
-\frac{p-1}{2}p^{e+m+s-3}, &  Tr(\gamma^{p^l+1}) \in C_0^{(2,p)},\\
\frac{p+1}{2}p^{e+m+s-3}, &  Tr(\gamma^{p^l+1}) \in C_1^{(2,p)},
\end{cases}
\quad \text{where}~i=0,1.
\end{equation*}
\end{lemma}
\textbf{Proof:} Similar to the proof of Lemma \ref{W3} and Lemma \ref{W5}, by utilizing Lemma \ref{weil sums2}, Lemma \ref{quadratic sums}, the definition of quadratic Gauss sums, Lemma \ref{quadratic character} and Lemma \ref{new sums}, we can also get the desired conclusion, and we will not demonstrate it in detail here.\hfill$\square$

Based on the discussion above, substituting $\psi_{i1}$, $\psi_{i2}$, $\psi_{i3}$, $\psi_{i4}$ to Eq.(\ref{5}), we can get the values of $T_i$\,$(i=0,1)$.
\begin{lemma}\label{T1}
With the notations above and $m/s \equiv 1\,mod\,2$, we have
\begin{enumerate}
  \item If $a = 0, b = 0$, then $T_i=n_i$,
  \item If $a = 0, b \neq 0$,
  \begin{enumerate}
    \item[(1)] $b \in \mathbb{F}_{p}^*$, then $T_i=0$,
    \item[(2)] $b \in \mathbb{F}_{q}^*/\mathbb{F}_{p}^*$, then
    $T_i=\frac{p-1}{2}(p^{2e-3}+p^{e+m-3})$,
  \end{enumerate}
  \item If $a \neq 0, b = 0$,
  \begin{enumerate}
\item[(1)] If the equation $X^{p^{2l}}+X=-a^{p^l}$ has a unique solution $\gamma$ over $\mathbb{F}_{q}$, and $Tr(\gamma^{p^l+1})=0$, then $$T_i=\frac{p-1}{2}(p^{2e-3}+p^{e+m-2}),$$
\item[(2)] If the equation $X^{p^{2l}}+X=-a^{p^l}$ has a unique solution $\gamma$ over $\mathbb{F}_{q}$, and $Tr(\gamma^{p^l+1}) \neq 0$, then
      \begin{equation*}
      T_i=
      \begin{cases}
      \frac{p-1}{2}(p^{2e-3}+p^{e+m-2}), & Tr(\gamma^{p^l+1}) \in C_0^{(2,p)}, \\
      \frac{p-1}{2}(p^{2e-3}-p^{e+m-2}), & Tr(\gamma^{p^l+1}) \in C_1^{(2,p)},
    \end{cases}
    \end{equation*}
\end{enumerate}
\item If $a\neq0 , b\neq0$
\begin{enumerate}
\item[(1)] $b\in\mathbb{F}_p^*,~Tr(\gamma^{p^l+1})\in C_0^{(2,p)},~\text{or } Tr(\gamma^{p^l+1})=0$, then $$T_i=\frac{p-1}{2}p^{2e-3},$$
\item[(2)]  $b\in\mathbb{F}_p^*,~Tr(\gamma^{p^l+1})\in C_1^{(2,p)}$, then
 \begin{equation*}
  T_i=\frac{p-1}{2}p^{2e-3}+p^{e+m-2},
 \end{equation*}
\item[(3)] $b\in\mathbb{F}_q^*\backslash\mathbb{F}_p^*$, then ~$T_i=\frac{p-1}{2}(p^{2e-3}+p^{e+m-3})$,
\end{enumerate}
\end{enumerate}
where $i=0,1$.
\end{lemma}
\textbf{Proof:} We only prove the case 3(2) here since the proofs of the other cases are very similar.\\
\indent In this case, $a \ne 0$, $b=0$, and $Tr(\gamma^{p^l+1}) \ne 0$,
\begin{enumerate}
\item [(1)] $\psi_{i1}=0$,
\item [(2)] $\psi_{i2}=0$,
\item [(3)] $\psi_{i3}=-\frac{p-1}{2}p^{e+m-2}\eta\big(-Tr(\gamma^{p^l+1})\big)-\frac{p-1}{2}p^{e+m-3}$,
\item [(4)] $\psi_{i4}=0$.
\end{enumerate}
Applying these values into Eq.(\ref{5}), we can get
\begin{align*}
T_i&=\frac{p-1}{2}p^{2e-3}+\frac{p-1}{2}p^{e+m-2}\eta(Tr(\gamma^{{p^l}+1}))\\&=
\begin{cases}
\frac{p-1}{2}(p^{2e-3}+p^{e+m-2}), & Tr(\gamma^{p^l+1}) \in C_0^{(2,p)}, \\
\frac{p-1}{2}(p^{2e-3}-p^{e+m-2}), & Tr(\gamma^{p^l+1}) \in C_1^{(2,p)},
\end{cases}
\quad \text{where}~i=0,1.
\end{align*}
Other cases are similar to the above solving process, which will not be repeated here. Then we can get the desired conclusion.\hfill $\square$

\begin{lemma}\label{T2}
With the notations above and $m/s \equiv 0\,mod\,2$, we have
\begin{enumerate}
  \item If $a = 0, b = 0$, then $T_i=n_i$,
  \item If $a = 0, b \neq 0$,
  \begin{enumerate}
    \item[(1)] $b \in \mathbb{F}_{p}^*$, then $T_i=0$,
    \item[(2)] $b \in \mathbb{F}_{q}^*/\mathbb{F}_{p}^*$, then
    $T_i=\frac{p-1}{2}(p^{2e-3}+p^{e+m+s-3})$,
  \end{enumerate}
  \item If $a \neq 0, b = 0$, then
  \begin{enumerate}
    \item[(1)] If the equation $X^{p^{2l}}+X=-a^{p^l}$ has no solution in $\mathbb{F}_{q}$, then $$T_i=\frac{p-1}{2}(p^{2e-3}+p^{e+m+s-3}),$$
    \item[(2)] If the equation $X^{p^{2l}}+X=-a^{p^l}$ has some solution $\gamma$ in $\mathbb{F}_{q}$, $Tr(\gamma^{p^l+1})=0$, then $$T_i=\frac{p-1}{2}(p^{2e-3}+p^{e+m+s-2}),$$
    \item[(3)] If the equation $X^{p^{2l}}+X=-a^{p^l}$ has some solution $\gamma$ in $\mathbb{F}_{q}$, $Tr(\gamma^{p^l+1}) \neq 0$, then
      \begin{equation*}
      T_i=
      \begin{cases}
      \frac{p-1}{2}(p^{2e-3}+p^{e+m+s-2}), & Tr(\gamma^{p^l+1}) \in C_0^{(2,p)}, \\
      \frac{p-1}{2}(p^{2e-3}-p^{e+m+s-2}), & Tr(\gamma^{p^l+1}) \in C_1^{(2,p)},
    \end{cases}
    \end{equation*}
  \end{enumerate}

  \item If $a \neq 0, b \neq 0$,
  \begin{enumerate}
    \item[(1)]  $b \in \mathbb{F}_{p}^*$,
    \begin{enumerate}
      \item[(i)] If the equation $X^{p^{2l}}+X=-a^{p^l}$ has no solution in $\mathbb{F}_{q}$, then $$T_i=\frac{p-1}{2}(p^{2e-3}+p^{e+m+s-3}),$$
      \item[(ii)] If the equation $X^{p^{2l}}+X=-a^{p^l}$ has some solution $\gamma$ in $\mathbb{F}_{q}$, and $Tr(\gamma^{p^l+1})=0$, then
          $$T_i=\frac{p-1}{2}p^{2e-3},$$
      \item[(iii)] If the equation $X^{p^{2l}}+X=-a^{p^l}$ has some solution $\gamma$ in $\mathbb{F}_{q}$, and  $Tr(\gamma^{p^l+1}) \neq 0$, then
      \begin{equation*}
      T_i=
      \begin{cases}
      \frac{p-1}{2}p^{2e-3}, & Tr(\gamma^{p^l+1}) \in C_0^{(2,p)}, \\
      \frac{p-1}{2}p^{2e-3}+p^{e+m+s-2}, & Tr(\gamma^{p^l+1}) \in C_1^{(2,p)},
    \end{cases}
    \end{equation*}
    \end{enumerate}
    \item[(2)] $b \in \mathbb{F}_{q}^*/\mathbb{F}_{p}^*$, then $$T_i=\frac{p-1}{2}(p^{2e-3}+p^{e+m+s-3}),$$
  \end{enumerate}
\end{enumerate}
where $i=0,1$.
\end{lemma}
\textbf{Proof:} Like Lemma \ref{T1}, we only prove the case 4(1)(iii) here. \\
\indent In this case, $a \ne 0$, $b \ne 0$, $b \in \mathbb{F}_p^*$, and $Tr(\gamma^{p^l+1}) \ne 0$,
\begin{enumerate}
\item [(1)] $\psi_{i1}=0$,
\item [(2)] $\psi_{i2}=0$,
\item [(3)] $\psi_{i3}=0$,
\item [(4)]
\begin{equation*}
\psi_{i4}=
\begin{cases}
-\frac{p-1}{2}p^{e+m+s-3}, & Tr(\gamma^{p^l+1}) \in C_0^{(2,p)}, \\
\frac{p+1}{2}p^{e+m+s-3}, & Tr(\gamma^{p^l+1}) \in C_1^{(2,p)}.
\end{cases}
\end{equation*}
\end{enumerate}
Substituting the above evaluations into Eq.(\ref{5}), we can get
\begin{equation*}
T_i=
\begin{cases}
\frac{p-1}{2}p^{2e-3}, & Tr(\gamma^{p^l+1}) \in C_0^{(2,p)}, \\
\frac{p-1}{2}p^{2e-3}+p^{e+m+s-2}, & Tr(\gamma^{p^l+1}) \in C_1^{(2,p)},
\end{cases}
\quad \text{where}~i=0,1.
\end{equation*}
\indent By a similar argument as that of Lemma \ref{T1}, we can get the desired conclusions. So we will not discuss it any more.\hfill $\square$

\textbf{Proof of Theorem \ref{weight1}:} With Lemma \ref{N(u,v)}, Lemma \ref{T1} and Eq.(\ref{3}), we can get the lengths and the weights of the codewords, which are shown in Table 1. We denote the non-zero weights of the lines $1-6$ in Table \ref{1} by $wt_{ij}$, and the corresponding multiplicity by $A_{wt_{ij}},\,1 \leq j \leq 6$\,($i=0,1$).
\begin{enumerate}
\item [(1)] If $a=b=0$, ~$wt_{i0}=0$,
\item [(2)] If $a=0$, $b \ne 0$, $b \in \mathbb{F}_p^*$, $wt_{i1}=n_i=\frac{p-1}{2}(p^{2e-2}+p^{e+m-2})$,
             \begin{align*}
             A_{wt_{i1}}&=|\{(a,b): a=0, b \ne 0, b \in \mathbb{F}_p^*\}|\\
                       &=|\{(a,b): a=0, b \in \mathbb{F}_p^*\}|\\
                       &=p-1,
            \end{align*}
\item [(3)] If $a=0$, $b \ne 0$, $b \in \mathbb{F}_q^*\backslash\mathbb{F}_p^*$, ~or~$a \ne 0$,\,$b \ne 0$,\,$b \in \mathbb{F}_q^*\backslash\mathbb{F}_p^*$,
             \begin{align*}
             wt_{i2}&=\frac{(p-1)^2}{2}(p^{2e-3}+p^{e+m-3})\\
             A_{wt_{i2}}&=|\{(a,b): a=0, b \ne 0, b \in \mathbb{F}_q^*\backslash\mathbb{F}_p^*, ~\text{or }~a \ne 0, b \ne 0, b \in \mathbb{F}_q^*\backslash\mathbb{F}_p^*\}|\\
             &=|\{(a,b): a \in \mathbb{F}_q, b \in \mathbb{F}_q^*\backslash\mathbb{F}_p^*\}|\\
             &=q[(q-1)-(p-1)]\\
             &=q(q-p)\\
             &=p^e(p^e-p),
             \end{align*}
\item [(4)] If $a \ne 0$, $b=0$, $Tr(\gamma^{p^l+1})=0$ or $Tr(\gamma^{p^l+1}) \in C_0^{(2,p)}$, $wt_{i3}=\displaystyle\frac{(p-1)^2}{2}p^{2e-3}$,
\item [(5)] If $a \ne 0$, $b=0$, $Tr(\gamma^{p^l+1}) \in C_1^{(2,p)}$, $$wt_{i4}=\frac{(p-1)^2}{2}p^{2e-3}+(p-1)p^{e+m-2},$$
\item [(6)] If $a \ne 0$, $b \ne 0$, $b \in \mathbb{F}_p^*$, $Tr(\gamma^{p^l+1})=0$ or $Tr(\gamma^{p^l+1}) \in C_0^{(2,p)}$, $$wt_{i5}=\frac{(p-1)^2}{2}p^{2e-3}+\frac{p-1}{2}p^{e+m-2},$$
\item [(7)] If $a \ne 0$, $b \ne 0$, $b \in \mathbb{F}_p^*$, $Tr(\gamma^{p^l+1}) \in C_1^{(2,p)}$, $$wt_{i6}=\frac{(p-1)^2}{2}p^{2e-3}+\frac{p-3}{2}p^{e+m-2}.$$
\end{enumerate}
Let
\begin{align*}
A_{wt_{i3}}&=|\{(a,b): a \ne 0, b=0, Tr(\gamma^{p^l+1})=0, ~\text{or } Tr(\gamma^{p^l+1}) \in C_0^{(2,p)}\}|=x,\\
A_{wt_{i4}}&=|\{(a,b): a \ne 0, b=0, Tr(\gamma^{p^l+1}) \in C_1^{(2,p)}\}|=y.
\end{align*}
Then we can have
\begin{align*}
A_{wt_{i5}}&=|\{(a,b): a \ne 0, b \ne 0, b \in \mathbb{F}_p^*, Tr(\gamma^{p^l+1})=0, \text{or } Tr(\gamma^{p^l+1}) \in C_0^{(2,p)}\}|=(p-1)x,\\
A_{wt_{i6}}&=|\{(a,b): a \ne 0, b \ne 0, b \in \mathbb{F}_p^*, Tr(\gamma^{p^l+1}) \in C_1^{(2,p)}\}|=(p-1)y.
\end{align*}
By the first three Pless Power Moments \cite{HP}, we have
\begin{eqnarray*}
&&\sum_{j=1}^6 {A_{wt_{ij}}}=p^{2e}-1,\qquad \qquad \qquad \qquad \qquad \quad \\
&&\sum_{j=1}^6 {wt_{ij}}A_{wt_{ij}}=p^{2e-1}(p-1)n_i, \\
&&\sum_{j=1}^6 {wt_{ij}}^2A_{wt_{ij}}= p^{2e-2}(p-1)n_i(pn_i-n_i+1).
\end{eqnarray*}
Using Maple 18, we can get the values of $A_{wt_{i3}}$-$A_{wt_{i6}}$, which are given in Table 1.
Thus we can get the desired conclusions presented in Theorem \ref{weight1}, and complete the proof.\hfill $\square$

\textbf{Proof of Theorem \ref{weight2}:} With Lemma \ref{N(u,v)}, Lemma \ref{T2} and Eq.(\ref{3}), we can also get the lengths and the weights of the codewords, which are shown in Table 2. We denote the non-zero weights of the lines $1-6$ in Table \ref{2} by $wt_{ij}$, and the corresponding multiplicity by $A_{wt_{ij}},\,1 \leq j \leq 6$\,($i=0,1$).

We find that if $a\in\mathbb{F}_q$ and $b\in\mathbb{F}_q^*\backslash\mathbb{F}_p^*$, $wt_{i2}=\frac{(p-1)^2}{2}(p^{2e-3}+p^{e+m+s-3})$, and the multiplicity $A_{wt_{i21}}=p^{e}(p^{e}-p);$ if $b \in \mathbb{F}_p$ and $X^{p^{2l}}+X=-a^{p^{l}}$ has no solution in $\mathbb{F}_q$, by Lemma \ref{weil sums4}, the multiplicity $A_{wt_{i22}}=p(p^e-p^{e-2s})$; thus $A_{wt_{i2}}=A_{wt_{i21}}+A_{wt_{i22}}=p^e(p^e-p^{1-2s})$.

The rest of the proof process is similar to that of Theorem \ref{weight1}, and will not be repeated here.\hfill $\square$

\section{Concluding remarks}\label{section-5}\rm
In this paper, inspired by the work in \cite{XXRF}, two classes of five-weight or six-weight linear codes were constructed with their weight enumerators settled using Weil sums and a new type of exponential sums. At the same time, some optimal or almost optimal linear code was found. It would be nice if more linear codes with few weights can be presented.


\end{document}